\documentclass[notitlepage,natbib,aps,prd,twocolumn,amsmath,amsfonts,nofootinbib,preprintnumbers,superscriptaddress]{revtex4-1}
\pdfoutput=1
\usepackage{amssymb,amsmath,latexsym,mathrsfs}
\usepackage{url}
\usepackage{enumitem}
\usepackage{graphicx}
\usepackage{booktabs}
\usepackage{fontawesome}
\usepackage[usenames,dvipsnames]{color}
\usepackage[breaklinks,colorlinks,urlcolor=magenta,citecolor=magenta,linkcolor=magenta]{hyperref}
\usepackage{multirow}
\usepackage{float}
\usepackage{cases}
\usepackage{blindtext}
\usepackage{pifont}
\usepackage{hhline}
\usepackage{xcolor}
\usepackage[normalem]{ulem}
\definecolor{linkcolor}{rgb}{0.0, 0.47, 0.75}
\definecolor{citecolor}{rgb}{1.0, 0.5, 0.0}
\hypersetup{
  linkcolor  = linkcolor,
  citecolor  = linkcolor,
  urlcolor   = linkcolor,
  colorlinks = true
}

\makeatletter
\def\maketitle{
\@author@finish
\title@column\titleblock@produce
\suppressfloats[t]}
\makeatother

\begin{document}

\preprint{CERN-TH-2024-127}
\vspace{5mm}

\title{Leveraging Time-Dependent Instrumental Noise for LISA SGWB Analysis}

\author{James Alvey}
\email{j.b.g.alvey@uva.nl}
\thanks{ORCID: \href{https://orcid.org/0000-0003-2020-0803}{0000-0003-2020-0803}}
\affiliation{GRAPPA Institute, Institute for Theoretical Physics Amsterdam,\\
University of Amsterdam, Science Park 904, 1098 XH Amsterdam, The Netherlands}

\author{Uddipta Bhardwaj}
\email{u.bhardwaj@uva.nl}
\thanks{ORCID: \href{https://orcid.org/0000-0003-1233-4174}{0000-0003-1233-4174}}
\affiliation{GRAPPA Institute, Anton Pannekoek Institute for Astronomy and Institute of High-Energy Physics,\\
University of Amsterdam, Science Park 904, 1098 XH Amsterdam, The Netherlands}

\author{Valerie Domcke}
\email{valerie.domcke@cern.ch}
\thanks{ORCID: \href{https://orcid.org/0000-0002-7208-4464}{0000-0002-7208-4464}}
\affiliation{Theoretical Physics Department, CERN, 1211 Geneva 23, Switzerland }

\author{Mauro Pieroni}
\email{mauro.pieroni@cern.ch}
\thanks{ORCID: \href{https://orcid.org/0000-0003-0665-266X}{0000-0003-0665-266X}}
\affiliation{Theoretical Physics Department, CERN, 1211 Geneva 23, Switzerland }

\author{Christoph Weniger}
\email{c.weniger@uva.nl}
\thanks{ORCID: \href{https://orcid.org/0000-0001-7579-8684}{0000-0001-7579-8684}}
\affiliation{GRAPPA Institute, Institute for Theoretical Physics Amsterdam,\\
University of Amsterdam, Science Park 904, 1098 XH Amsterdam, The Netherlands}

\begin{abstract}
\noindent Variations in the instrumental noise of the Laser Interferometer Space Antenna (LISA) over time are expected as a result of \emph{e.g.}~scheduled satellite operations or unscheduled glitches. We demonstrate that these fluctuations can be leveraged to improve the sensitivity to stochastic gravitational wave backgrounds (SGWBs) compared to the stationary noise scenario. This requires optimal use of data segments with downward noise fluctuations, and thus a data analysis pipeline capable of analysing and combining shorter time segments of mission data. We propose that simulation based inference is well suited for this challenge. In an approximate, but state-of-the-art, modeling setup, we show by comparison with Fisher Information Matrix estimates that the optimal information gain can be achieved in practice.

\vspace*{5pt} \noindent \faGithub \,\textbf{\texttt{GitHub}}: The \texttt{saqqara} simulation and inference library is available \href{https://github.com/peregrine-gw/saqqara}{here} (\texttt{peregrine-gw/saqqara}). In addition, the TMNRE implementation \texttt{swyft} is available \href{https://github.com/undark-lab/swyft}{here} (\texttt{undark-lab/swyft}). Finally, the \texttt{GW\textunderscore response} code to compute the LISA response function is available \href{https://github.com/Mauropieroni/GW_response}{here} (\texttt{Mauropieroni/gw\textunderscore response})
\end{abstract}

\maketitle
\hypersetup{
  linkcolor  = linkcolor,
  citecolor  = linkcolor,
  urlcolor   = linkcolor
}

\section{Introduction}\label{sec:intro}
\noindent The next generation of gravitational wave (GW) interferometers, in particular the Laser Interferometer Space Antenna (LISA)~\cite{LISA:2017pwj}, the Einstein Telescope (ET)~\cite{Punturo:2010zz}, and Cosmic Explorer (CE)~\cite{Reitze:2019iox} will revolutionize our ability to explore the Universe through GWs. A key to this is the extension of the frequency coverage as well as the tremendous increase in strain sensitivity compared to current GW detectors~\cite{LIGOScientific:2016aoc,VIRGO:2014yos, KAGRA:2020tym}. However, there is a significant data analysis challenge that comes with this increased sensitivity. While current ground-based GW detectors are noise dominated with sparse signals due to merging compact objects, the next generation detectors will be signal dominated, with thousands of resolvable binary systems as well as astrophysical (and possibly cosmological) stochastic gravitational wave backgrounds (SGWBs) within instrument reach. This calls for a paradigm change in the data analysis~\cite{Cornish:2005qw,Littenberg:2023xpl}. 

In this paper, we focus on the particular challenge of detecting an SGWB in LISA under these circumstances. In the LISA band, we expect astrophysical SGWBs due to unresolved (\emph{i.e.} faint) binaries of compact objects (neutron stars, white dwarfs, black holes). Moreover, cosmological SGWBs are predicted in many extensions of the Standard Model of particle physics, due to, \emph{e.g.}, first order phase transitions, topological defects, the formation of primordial black holes, or in some models of cosmic inflation, see \emph{e.g.}, Ref.~\cite{Caprini:2018mtu} for an overview. Thus, observing or constraining the presence of such SGWBs provides a unique window to the early universe and unexplored particle physics regimes.

In rather simplified terms, the challenge in searching for such signals in LISA is twofold. Firstly, any fit to the data needs to account for the presence of GW signals from loud binaries, the possible presence of SGWBs, and the instrument noise. This is at the core of the `global fit' program~\cite{Cornish:2005qw, Littenberg:2023xpl, Strub:2024kbe, Katz:2024oqg} currently based on Monte Carlo Markov Chain (MCMC) techniques, although the first steps towards implementing this using simulation based inference (SBI, see Ref.~\cite{Cranmer:2019eaq} for a review) have been taken in Refs.~\cite{Alvey:2023npw,Dimitriou:2023knw,Korsakova:2024sut}. Secondly, disentangling an SGWB contribution from the instrument noise suffers from the limited knowledge of the latter. This situation is exacerbated by (i) the lack of `down-time' in the detector, \emph{i.e.}, the lack of a signal free environment to measure the noise properties, (ii) the absence of independent data channels (as in a network of ground-based detectors) to cross-correlate with and suppress the noise, and (iii) the multitude of possible signal templates. Given this, existing analyses have relied on fixed templates for the noise~\cite{Karnesis:2019mph, Caprini:2019pxz, Pieroni:2020rob, Flauger:2020qyi}, the signal~\cite{Baghi:2023qnq, Muratore:2023gxh} or both~\cite{Boileau:2020rpg, Hartwig:2023pft}. See also~\cite{Pozzoli:2023lgz}, for an analysis with flexible signal and noise templates.

An additional simplification used in all these approaches is to assume that the SGWB (in the appropriate reference frame) and the noise are stationary, \emph{i.e.}, the data can be modelled by drawing different realizations from fixed, time-independent distributions. While for SGWBs this is fully justified on the time-scales of relevance for LISA, the instrumental noise is known to suffer from scheduled (due to instrument operation) and unscheduled glitches, \emph{i.e.}, interruptions in the data stream~\cite{Baghi:2021tfd,Robson:2018jly,Seoane:2021kkk,talkGair}\footnote{See also~\cite{Robson:2018jly} (LISA) and \cite{Pankow:2018qpo,Cornish:2014kda} (LIGO) for work on distinguishing glitches from transient signals.}. After such events, in particular if due to a change in the instrument configuration, there is no reason to expect the noise level to be exactly the same before and after. Our goal in this paper is twofold: (i) to explore the implications of time-varying instrumental noise on the SGWB reconstruction, and (ii) to demonstrate that SBI techniques are well-suited for analysing this scenario.

To this end, we have developed public codes (see \href{https://github.com/Mauropieroni/GW_response}{\texttt{Mauropieroni/gw\textunderscore response}} and \href{https://github.com/peregrine-gw/saqqara}{\texttt{peregrine-gw/saqqara}}) to simulate and analyze LISA data. The simulator features a modular structure for the signal components, the noise components, and the LISA response functions. In particular, compared to earlier work~\cite{Alvey:2023npw}, we include all three data channels of LISA and account for the full $\sim 3$ years of effective mission duration~\cite{Colpi:2024xhw,Seoane:2021kkk}. In this paper, we are in particular interested in the impact of more realistic noise modeling on the ability to reconstruct SGWB parameters. The base LISA model is a two-parameter noise model parametrizing the amplitude of the test mass (TM) and optical metrology system (OMS) noise assuming a given spectral shape. Being more realistic, however, the noise levels should be expected to vary in time, nominally lying between the current best estimate and the allocated noise budget~\cite{Colpi:2024xhw}. To investigate the impact of non-stationary noise on the ability to search for SGWBs, we cut the data streams into segments of 11.5 days (motivated by the cadence of scheduled satellite operations, although we note that any segment length could in principle be used here) where we assume the noise to be well-described by the two-parameter base LISA noise model. We then draw independent fiducial values for the two noise parameters in each segment from Gaussian distributions motivated by the allocated noise budget~\cite{Colpi:2024xhw}. 

This setup is, of course, still rather simplified compared to the full LISA data analysis challenge. However, the modular setup of our code lays the groundwork for several possible extensions, such as further refining the underlying (parametric) noise model, including different signal templates, or adding GW signals from different source populations of binaries.

The rest of this work is structured as follows: in Sec.~\ref{sec:modeling}, we start by describing the properties of the different modeling components included in our analysis. We proceed, in Sec.~\ref{sec:fisher}, by using the Fisher Information Matrix (FIM) formalism to demonstrate that with a suitable analysis strategy, we can (in principle) leverage the rare low-noise data segments to increase the signal-to-noise ratio (SNR) of the signal. With this expectation, in Sec.~\ref{sec:sbi}, we show that our SBI pipeline can efficiently achieve close to the optimal information gain in this scenario. Ultimately this highlights that time variation in the noise levels can allow for a more accurate reconstruction of the SGWB parameters compared to a constant noise scenario with fiducial values for the noise parameters taken to be the mean values of the distribution. Finally, in Sec.~\ref{sec:conclusions}, we conclude and discuss future perspectives for our work.

\section{Signal, Instrumental and Time-varying Noise Modeling}\label{sec:modeling}
\noindent Taking a step towards more realistic modeling of the instrument noise, our goal is to investigate the ability to search for SGWBs in the presence of non-stationary noise components. The naive expectation may be that this added complication could lead to a degradation of our ability to reconstruct the signal parameters. However, as we demonstrate below using both FIM forecasts and SBI techniques, this is not necessarily the case.  If the data analysis method succeeds in leveraging the gain in the SNR from the low noise segments, this can overcompensate for the loss in information in the high noise segments. Concretely, we will demonstrate that for a sufficiently large number of data segments, the signal parameter reconstruction with time varying noise can be more precise than for a constant noise level with mean amplitude.

In this section, we summarize the main ingredients used in the analysis regarding the data modeling. We start by introducing the instrumental noise model employed in our analysis and Time Delay Interferometry (TDI). We proceed by describing the instrumental response function and conclude by providing the template we use for the SGWB. The discussion and notation closely follow the one employed in~\cite{Hartwig:2023pft} to which we refer the reader for further details.

Before discussing these specific topics, we first introduce the conventions used throughout the paper. We use lower Roman letters ($i, j, \ldots$) to index the three LISA satellites labeled as 123. We denote with $\eta_{ij}(t)$ the measurement performed at time $t$ on the ``link'' $ij$, corresponding to the path connecting satellite $j$ to satellite $i$. Six links with different time-dependent lengths $L_{ij}(t)$ form the LISA configuration\footnote{Notice that since LISA rotates around its center of mass, even in the absence of GWs, the travel time for a photon leaving satellite $j$ at time $(t - L_{ij} / c)$, reaching satellite $i$ at time $t$ is different to the one for a photon traveling in the opposite direction from satellite $i$ to satellite $j$, \emph{i.e.}, $L_{ij} \neq L_{ji}$. This is known as the Sagnac effect.}. The measurement is defined as the time derivative of the difference between the measured travel time, including noise and, if present, GWs, and the travel time in the absence of such effects. In the following, we denote with $\eta_{ij}^{\rm GW}(t)$ and $\eta_{ij}^{\rm N}(t)$ the GW and noise contributions to the single link measurements, respectively. While in the analysis performed in Secs.~\ref{sec:fisher} and \ref{sec:sbi}, we will restrict ourselves to the case where $L_{ij} (t) = L = 2.5$ million kilometers (\emph{i.e.}, a static equilateral configuration) and uniform noise parameters across all satellites (albeit with independent noise realizations), in this section we provide the general formulae that do not rely on this assumption. Finally, $D_{ij}$ denotes the delay operator, whose action on a single link measurement $\eta_{lm}(t)$, under the assumptions stated above can be expressed as $D_{ij} \eta_{lm}(t) = \eta_{lm}(t - L_{ij} / c)$. The simulator defined in \href{https://github.com/peregrine-gw/saqqara}{\texttt{peregrine-gw/saqqara}} and the \href{https://github.com/Mauropieroni/GW_response}{\texttt{GW\textunderscore response}} code closely follows the notation and procedure described in the remainder of this section. Looking forwards, this setup also allows for a straightforward implementation of more complex noise, signal, and instrument models beyond the specific choices used in this paper. 

As a final comment, for consistency with most literature, in this section, we work with both positive and negative frequencies. For computational reasons, however, in all our codes we work with positive frequencies only. For this reason, all equations in the other sections of this paper (including App.~\ref{app:fisher}), are written in terms of positive frequencies only. 

\subsection{Time Delay Interferometry}
\label{sec:TDI}
\noindent Time Delay Interferometry, or TDI~\cite{Tinto:2020fcc} is a data processing technique that consists of combining measurements performed at different times to achieve noise suppression. In LISA, this technique will be crucial to reduce laser frequency noise, which otherwise dominates the noise budget, jeopardizing the prospects of any GW detection~\cite{LISA:2017pwj,LISA_performance}. While it is possible to define several TDI variables~\cite{Armstrong_1999, Prince:2002hp, Shaddock:2003bc, Shaddock:2003dj, Tinto:2003vj, Vallisneri:2005ji, Muratore:2020mdf, Muratore:2021uqj}, in this work we restrict our focus to two TDI bases: the Michelson variables, typically dubbed XYZ, and the diagonalized version, denoted by AET~\cite{Prince:2002hp}. In the first-generation version\footnote{First-generation TDI variables ensure laser noise suppression only in the limit of constant arm lengths, neglecting the Sagnac effect $L_{ij} \neq L_{ji}$. If these assumptions are relaxed, more complicated TDI variables are required (see, \emph{e.g.}, Ref.~\cite{Tinto:2003vj}).}, the TDI X variable, describing the difference in time delay between a return trip along the 12-arm and the 13-arm, is defined as
\begin{multline}
    \label{eq:TDI_X}
	{\rm X}  = (1 - D_{13}D_{31})(\eta_{12} + D_{12} \eta_{21}) \\ - (1 - D_{12}D_{21})(\eta_{13} + D_{13} \eta_{31}) \; .
\end{multline}
The $(1 - D_{1x}D_{x1})$ factors in this expression ensure that, in the absence of GWs, the total travel time in the two directions is the same for unequal (but constant) arm lengths. The Y and Z variables are defined through cyclic permutations of the indices appearing in this equation. Starting from the XYZ variables, the AET basis is then defined as
\begin{equation}
	{\rm A} = \frac{{\rm Z} - {\rm X}}{\sqrt{2}}\;, \quad  {\rm E} = \frac{{\rm X} - 2 {\rm Y} + {\rm Z}}{\sqrt{6}} \;, \quad  {\rm T}=\frac{{\rm X} + {\rm Y} + {\rm Z}}{\sqrt{3}}  \; .
\end{equation}
For equal arm lengths and equal noise levels across the three satellites, this basis removes the cross-correlations between different channels both for signal and noise. Moreover, it has the property that the T channel is nearly signal insensitive at low frequencies~\cite{Prince:2002hp} (hence it is known as a null channel), making it a good monitor for some noise components.

Notice that the XYZ, and similarly the AET variables, are defined as linear combinations of the single-link variables. As a consequence, the mapping from the single-link variables to any TDI basis (including the relevant time delays) can be defined through a 3 by 6 matrix, say $\chi^I_{ij}$, where the index $I$ runs over the three variables in the TDI basis and $ij$ labels the 6 single link variables. Analogously, any Power or Cross Spectral Density (PSD/CSD) can be defined by applying a 3 by 3 by 6 by 6 tensor $\Lambda^{I,J}_{ij,lm} \equiv \chi^I_{ij} \chi^J_{lm}$ on the single-link PSDs and CSDs.

\subsection{LISA noise model} 
\label{sec:noise_model}
\noindent After suppressing the laser frequency noise using TDI techniques, the LISA noise budget is expected to be dominated by two components: the Test Mass (TM) acceleration noise, and the Optical Metrology System (OMS) noise~\cite{LISA_sciRD,LISA_performance,Colpi:2024xhw}. For this reason, the TM and OMS noises are typically known as ``secondary noises". Each satellite composing the LISA constellation contains two TMs, whose noises are, in principle, independent. Similarly, six independent OMS noise components will affect the LISA measurements. These noise components enter the single-link measurements as
\begin{equation}
	\eta_{ij}^\mathrm{N}(t) = n^\text{OMS}_{ij}(t) +  D_{ij} n_{ji}^\text{TM}(t) + n_{ij}^\text{TM}(t) \; .\label{eq:eta_noise}
\end{equation}
The CSDs of the TM and OMS noise can be formally expressed as
\begin{align}
    \langle \tilde n^\text{TM}_{ij}(f) \, \tilde n^\text{TM*}_{lm}(f') \rangle &= \frac{1}{2} \,  \delta_{ij,lm} \, S^\text{TM}_{ij}(f) \, \delta(f - f') \; ,  \label{eq:csd_noise} \\
	\langle \tilde n^\text{OMS}_{ij}(f) \, \tilde n^\text{OMS*}_{lm}(f') \rangle &= \frac{1}{2} \, \delta_{ij,lm} \, S^\text{OMS}_{ij}(f) \,\delta(f - f') , 	
\end{align}
where we assume the noise PSDs (in seconds) to be given by~\cite{LISA:2017pwj} 
\begin{multline}
	\label{eq:TM_OMS_noise_def}
	S_{ij}^\text{TM}(f) = A_{ij}^2  \; \times 7.737 \times 10^{-46} \times \left( \frac{L }{ 2.5 \times 10^9 \textrm{m} } \right)^2 \left(\frac{f_* }{f } \right)^2 \\ \times \left[1 + \left(\frac{ 0.4 \textrm{mHz}}{f}\right)^2 \right]  \left[1 + \left(\frac{f}{ 8 \textrm{mHz}} \right)^4 \right] \; \times \textrm{s} \;,
\end{multline}
\begin{multline}
	S_{ij}^\text{OMS}(f) = P_{ij}^2 \times 1.6 \times 10^{-43} \; \times \left(\frac{2.5 \times 10^9 \textrm{m} }{L} \right)^2 \;  \left(\frac{f }{f_* } \right)^2  \\ \times \left[1 + \left(\frac{2\textrm{mHz} }{f} \right)^4 \right] \;  \times \textrm{s}  \;,
\end{multline} 
where $ f_* \equiv (2 \pi L /c )^{-1} \simeq 19$ mHz is the characteristic frequency for LISA, and $A_{ij}$, $P_{ij}$ are the six TM and OMS dimensionless noise parameters. In the following, we will work with a simplified version of the noise model where $A_{ij} = A$, $P_{ij} = P$, with the fiducial noise levels specified in~\cite{LISA_sciRD} corresponding to $A_0 = 3$ and $P_0 = 15$. To model the time varying noise, for each data segment of 11.5 days, we draw the values of $A$ and $P$ from Gaussian distributions with mean $A_0$ and $P_0$ and standard deviation of $20\%$ of these fiducial values. This is motivated by the range between the allocated noise budget and the current best estimate in Fig.~7.1 of Ref.~\cite{Colpi:2024xhw}. Given these values of $A$ and $P$, we generate independently the noise realizations in each link and in each data segment.

\subsection{Instrument response function and signal model} 
\label{sec:response_function}
\noindent The signal contribution to the single-link measurement $\eta_{ij}^{\mathrm{GW}}(t)$ is the fractional frequency shift induced by GWs on the path of a photon traveling from satellite $j$ (at position $\vec x_j$) to satellite $i$ (at position $\vec x_i$). By expanding the GW signal in plane waves, this quantity can be expressed as
\begin{multline}
	\eta_{ij}^{\mathrm{GW}}(t) =  i \left( \frac{L_{ij}}{L} \right) \int_{-\infty}^{\infty} \textrm{d} f\,  \left(  \frac{f}{f_*}  \right)  \, \textrm{e}^{2\pi i f(t - L_{ij})} \\
	\int \textrm{d} \Omega_{\hat{k}} \, \textrm{e}^{-2\pi i f \hat{k}\cdot \vec x_i } \sum_\lambda \xi^\lambda_{ij}(f, \hat k) \, \tilde{h}_\lambda(f,\hat{k})  \; ,
\label{eq:etaij_final}
\end{multline}
where $\vec{k}$ is the GW wavevector, and $k$ is its norm, $\Omega_{\hat{k}}$ is the solid angle, $\tilde{h}_\lambda(f,\hat{k})$ are the Fourier coefficients with polarization index $\lambda$, and
\begin{multline}
	\xi^\lambda_{ij}\left(f, \hat k\right)= \mathrm{e}^{\pi i f L_{ij} ( 1 - \hat{k}\cdot \hat l_{ij})}  \; \mathrm{sinc}\left(\pi f L_{ij} ( 1 + \hat{k}\cdot \hat l_{ij})\right) \\ \;  \times \frac{\hat l^a_{ij}\hat l^b_{ij}}{2}e^\lambda_{ab}(\hat k) \; ,
\end{multline}
where $\hat l_{ij} = (\vec x_j - \vec x_i)/|\vec x_j - \vec x_i|$ is the unit vector going from $i$ to $j$ and $e^\lambda_{ab}(\hat k)$ are the GW polarization tensors defined as in~\cite{Hartwig:2023pft}. The single-link CSD (in the frequency domain) for the signal can then be expressed as
\begin{multline}
	S^{\mathrm{GW}}_{ij,mn}(f) \equiv \sum_\lambda \mathcal{R}^\lambda_{ij,mn} \, P_h^{\lambda \lambda}(f) \\\quad \equiv \left( \frac{L_{ij}}{L} \right) \left( \frac{L_{lm}}{L} \right) \left( \frac{f}{f_*} \right)^2 \sum_\lambda P_h^{\lambda \lambda }(f)  \hfill \\ \times \int \frac{\textrm{d} \Omega_{\hat{k}}}{4 \pi}  \; \textrm{e}^{-2\pi i f \hat{k}\cdot (\vec x_i - \vec x_m)}  \;  \xi^\lambda_{ij}(f, \hat k)  \, \xi^\lambda_{mn}(f, \hat k)^* \; ,
 \label{eq:CSDsignal}
\end{multline}
where we have introduced the (polarization-dependent) single-link response functions $\mathcal{R}^\lambda_{ij,mn} $ and substituted the statistical properties for a homogeneous and isotropic SGWB
\begin{equation} 
\label{eq:h-statistics}
	\langle \tilde h_\lambda(f,\hat k) \, \tilde h_{\lambda^\prime}^*(f', \hat k')\rangle = \delta(f - f')\delta(\hat k -\hat k')\delta_{ \lambda \lambda^{\prime} }\frac{P_{h}^{\lambda \lambda^{\prime}}(f)}{16\pi} \,. 
\end{equation}
Following the procedure discussed in Sec.~\ref{sec:TDI}, the single-link responses and signal CSDs can be suitably combined to get their analogues in any TDI basis. 

Note that, while we consider all noise components to be independent link-by-link there is only one SGWB realization at a given time, characterized by $\tilde{h}_\lambda(f, \hat{k})$ to be projected onto the different TDI variables. However, given the orthogonality of the AET basis, in practice, each channel measures statistically independent realizations of the sky-averaged signal. This is due to a re-weighting of contributions from the different angular directions by the corresponding sky-dependent response functions.

\subsection{Signal model}
\label{sec:signal_model}
\noindent Finally, we discuss our choice for the signal model. In this work, we restrict our study to one of the simplest possible choices, \emph{i.e.}, a power law template
\begin{align}\
 \Omega_\text{GW}(f) h^2 = 10^\alpha \, (f/f_*)^\gamma
\end{align}
with the two parameters $\alpha$ and $\gamma$ and the pivot scale $f_*$ set to $\sqrt{f_\mathrm{min} f_\mathrm{max}} \simeq 3.8 \times 10^{-3}\,\mathrm{Hz}$, where $f_\mathrm{min,max}$ are the minimum and maximum frequencies defined below in the case studies. This simple template serves as an illustrative example and allows for a thorough comparison between the different methods (FIM forecasts and SBI) presented here. As demonstrated in Refs.~\cite{Alvey:2023npw,Dimitriou:2023knw}, SBI techniques can be straightforwardly generalized to more complicated signal templates, including quasi-agnostic templates consisting of multiple broken power laws. The signal parameters are taken to be constant across all data segments.

In this work, we assume a static SGWB. An anisotropic SGWB would violate this assumption in the instrument reference frame,  due to the LISA antenna pattern sweeping the sky. However, this could be included in the modeling of the instrument response, thus recovering a time-independent SGWB~\cite{Digman:2022jmp}. 

Moreover, we ignore the presence of foreground signals such as inspirals and transient sources. See, however, Ref.~\cite{Alvey:2023npw} for a demonstration of an unbiased recovery of the SGWB signal parameters in the presence of mock foreground transients. In the context of this work, long duration foreground signals spanning more than one data segment may pose a particular challenge, as phase information is lost when the data segments are analyzed separately. However, for SGWB analyses, a possible solution may be to mask the corresponding frequencies of foreground inspirals. We leave a more detailed study of this problem to future work.

\section{Expectations: Fisher Forecasting}\label{sec:fisher}
\noindent In this section, we use FIM techniques (see below) to provide a benchmark estimate for the impact that time-varying noise modeling has on the reconstruction of an SGWB. For this purpose, we take the modeling setup described in the previous section and investigate the constraints on the model parameters $\alpha$ and $\gamma$ in the power law signal template across two case studies detailed below. We start this section by describing the case studies considered in this work. We proceed by introducing the FIM formalism and applying it to the case studies to forecast the accuracy in reconstructing the signal parameters in the different configurations. 

\subsection{Case Studies} 
\label{sec:cases}
\noindent In each case study, we consider the frequency-domain data $d_{i,\tau}^I = d^I_\tau(f_i)$ across the TDI channels $I = A, E, T$ split into $N_s = 100$ time segments of duration $T_s = 11.5 \, \mathrm{days}$, indexed by $\tau$ and defined on a frequency range with $f_\mathrm{min} = 3 \times 10^{-5} \, \mathrm{Hz}$ and $f_\mathrm{max} = 5 \times 10^{-1} \, \mathrm{Hz}$. In all cases, we take the fiducial values of the signal parameters to be $\alpha = -11$ and $\gamma = 0$. Consistently with App.~\ref{app:fisher}, we define the parameter vector $\theta = (\theta_h, \theta_{n,1}, \ldots, \theta_{n,N_s})$ to consist of a set of signal parameters $\theta_h = (\alpha, \gamma)$ and $N_s$ sets of noise parameters $\theta_{n,\tau} = (A_\tau, P_\tau)$ for each segment $\tau = 1...N_s$. In more detail, we consider:
\begin{itemize}
    \item \textbf{Case 1 (Fixed Noise):} For this benchmark case, we assume that the noise level does not vary across the various data segments\footnote{It can be shown either from an information perspective or directly at the level of the covariance matrix that this setup is equivalent (in the sense of constraints on $\alpha$ and $\gamma$) to analysing the full dataset as a single entity with only one set of noise parameters $(A_0, P_0) = (3, 15)$. This is not true for the more general \textbf{Case 2}.} so that the parameter vector $\theta$ is given by $\theta = (\alpha=-11, \gamma=0, A_1=3, P_1=15, \cdots, A_{N_s} = 3, P_{N_s} = 15)$. The FIM and the corresponding covariance matrix (see below), allows us to define a benchmark sensitivity for the signal parameters $\alpha$ and $\gamma$.
    
    \item \textbf{Case 2 (Time-varying Noise):} In the second case study, we investigate how time-varying noise across the various data segments impacts the determination of $\alpha$ and $\gamma$. Specifically, the parameter vector $\theta = (\alpha, \gamma, A_1, P_1, ...)$ that we evaluate the FIM on is now constructed as follows: fix $\alpha = -11$ and $\gamma = 0$ as per \textbf{Case 1}, then sample $N_s$ realizations of $(A_\tau \sim \mathcal{N}(3, 0.6), P_\tau \sim \mathcal{N}(15, 3))$ and populate the parameter vector. For each realization, the corresponding FIM is computed (which involves summing the individual FIMs across segments), inverted to obtain the covariance matrix, and the parameter sensitivities and correlations for $\alpha$ and $\gamma$ are read off from the relevant covariance matrix components. This is then repeated for $N_\mathrm{samples} = 5000$ realizations, generating a distribution over the expected sensitivities.
\end{itemize}

\noindent In \textbf{Case 2} the varying noise levels in each segment will inevitably lead to segments where the noise is larger than the corresponding fixed case \textbf{Case 1} (with $A = 3, P = 15$) and vice versa. Naturally, in those segments where the noise is larger than the average value, the constraints on the signal parameters will degrade. Likewise, if the noise is below average, the constraints on the signal parameters tighten. The question we are trying to answer in this section is how do these two effects balance out at the level of parameter estimation in our setup.

\subsection{Fisher Information Matrix Construction} 
\label{sec:fisher-construction}
\noindent In order to carry out these estimates, we construct the FIM $F_{AB}(\theta)$ for the parameters $\theta$ given our data modeling assumptions. The full derivation for the FIM over all data segments is given in App.~\ref{app:fisher}, and is implemented in the \texttt{sgwbfish} module of the \texttt{saqqara} code. In short, the FIM is given by the expectation $\mathbb{E}$ of the following quantity over data realizations $d$ under the data generation process $p(d | \theta)$,
\begin{equation}
    F_{AB}(\theta) = - \mathbb{E}_{d \sim p(d | \theta)} \left[\frac{\partial^2  \log p(d | \theta)}{\partial \theta_A \partial \theta_B} \right].
\end{equation}
where $A$ and $B$ run over the signal and noise parameters in all data segments, see App.~\ref{app:fisher} for details. After constructing the FIM, one can then compute the corresponding covariance matrix $C_{AB}(\theta) = F^{-1}_{AB}(\theta)$ by simply taking the matrix inverse. In our context, and for the results below, we are then interested in the relevant sub-matrix,  corresponding to the parameters of the signal template. These provide sensitivity estimates for the variance and covariance across the signal parameters, marginalised over the noise parameters. Specifically, we can consider the sub-matrix $C_h(\theta)$ of $C_{AB}(\theta)$ given by:
\begin{align}
    C_h(\theta = (\alpha, \gamma, A_1, P_1, \cdots)) &= \begin{bmatrix}
    C_{\alpha \alpha}(\theta) & C_{\alpha \gamma}(\theta) \\
    C_{\gamma \alpha}(\theta) & C_{\gamma \gamma}(\theta)
  \end{bmatrix} \\ \nonumber
  & \equiv \begin{bmatrix}
    \sigma_\alpha^2 & \sigma_{\alpha, \gamma} \\
    \sigma_{\alpha \gamma} & \sigma_\gamma^2
  \end{bmatrix},
\end{align}
where $\sigma_{\alpha}$, $\sigma_{\gamma}$ are the expected standard deviations on the reconstruction of the parameters $\alpha$ and $\gamma$ in the signal template, and $\sigma_{\alpha\gamma}$ is the covariance between them. For a given choice of $\theta$, these quantities directly provide estimates of the possible precision one can achieve when determining the model parameters across the case studies above.

\begin{figure*}
    \centering
    \includegraphics[width=\linewidth]{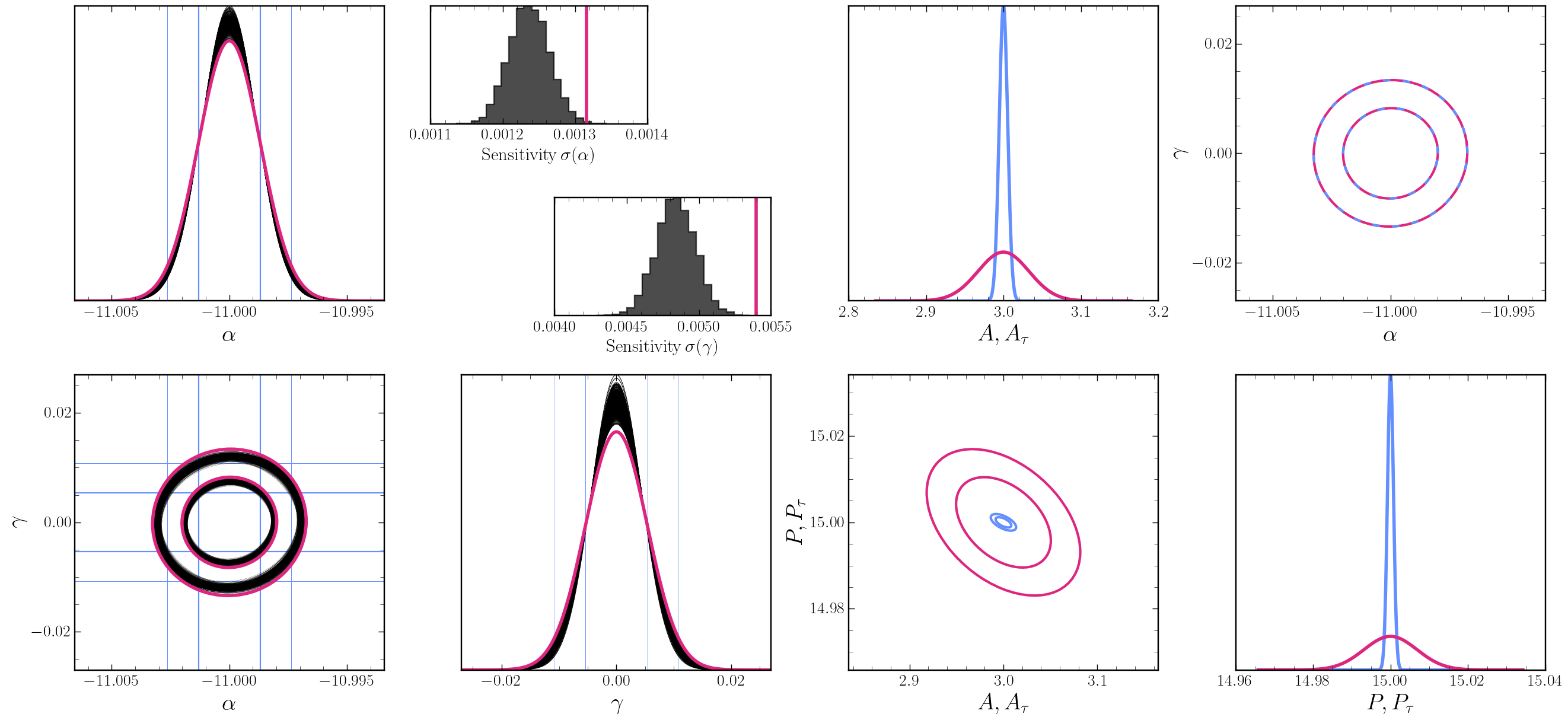} 
    \caption{\emph{Left:} FIM forecasts for the two-parameter signal model. The pink curves in the corner plot and the inset indicate the result for non-varying noise parameters. The black curves and the black histograms highlight the results for time-dependent noise parameters, sampled according to Sec.~\ref{sec:modeling}. The fiducial parameters taken here are given by: $\alpha = -11$, $\gamma = 0$, $A = 3$, $P = 15$, with $N_s = 100$ data segments, $f_\mathrm{min} = 3 \times 10^{-5} \, \mathrm{Hz}$, $f_\mathrm{max} = 5 \times 10^{-1}\,\mathrm{Hz}$, and $T_s = 11.5 \, \mathrm{days}$. \emph{Right:} Demonstration that in the case where the noise parameters are genuinely stationary across the dataset, we should expect the same sensitivity on the signal parameters $\alpha$ and $\gamma$ if we analyze the dataset assuming only one pair of noise parameters $(A = 3, P = 15)$, or allow them to vary across each segment $(A_\tau = 3, P_\tau = 15)$ for $\tau = 1...N_s$. Naturally, the constraints on the parameters covering the full dataset (blue curves/contours) ($A, P$) are tighter than the constraints on the parameters in any one segment $(A_\tau, P_\tau)$ (pink curves/contours). In the upper right panel, we show that nonetheless, the constraints and correlations on the signal parameters are identical to the left-hand scenario.}
    \label{fig:fisher}
\end{figure*}

\subsection{Fisher results} 
\label{sec:Fisher_results}
\noindent We carry out this analysis using the \texttt{sgwbfish} module in \texttt{saqqara} for the parameter choices provided above. The key results of this study are shown in Fig.~\ref{fig:fisher}. Specifically, the pink curves and vertical lines in this figure denote the results of \textbf{Case 1}. In the corner plots, we have generated the 1-dimensional Gaussians and the 2-dimensional ellipses using the covariance matrix $C_h(\theta)$. The results for \textbf{Case 2} are then shown via the black curves in the corner plots, and the black histograms in the upper panels. The histograms in particular show the distribution of the errors in the parameters relevant for signal reconstruction over realizations of $\theta$.

This leads to a very clear picture regarding the impact of the time varying noise on the sensitivity to SGWB. In particular, by analysing the data on a segment-by-segment basis, the increased sensitivity from the segments where the noise is reduced compared to the average outweighs the reverse impact. As such, we see that the distribution of errors on signal parameters lies below the benchmark value that would arise from either a) noise which maintains a fixed magnitude, or b) analysing the data assuming the noise remains fixed throughout, via some form of averaging process over segments. This motivates developing an analysis strategy that can leverage this improvement in performance induced by the low-noise segments of the data. We note that analysing the data allowing for time varying noise when in fact the noise levels are constant does not result in any information loss, \emph{i.e.}, the FIM results for the four parameter model ($\alpha, \gamma, A, P$), averaged over all segments, agree with those found in \textbf{Case 1}. This is demonstrated in the right-hand panel of Fig.~\ref{fig:fisher}. In the next section, we explore the extent to which SBI is a useful tool for this purpose and demonstrate that we can recover this expected sensitivity distribution in a full analysis of the data.

\section{Validation: Simulation Based Inference Framework}\label{sec:sbi}
\noindent In this section, we demonstrate that, consistently with the FIM results in Sec.~\ref{sec:fisher}, the SBI pipeline we have developed can leverage the information gain in the low-noise segments. This framework relies on analysing the data segment-by-segment, as stacking all data segments would instead result in data featuring the average noise level and the additional information will be lost. 
This showcases the benefit of an implicit likelihood method (here for the frequency coarse-grained single-segment data), for performing this analysis efficiently, as detailed below.
Below we describe our general strategy and technical implementation before presenting the results for the case study described above.

\subsection{SBI Analysis Strategy}

\noindent Ultimately, the possible benefits highlighted by the FIM analysis come down to analysing the data segment-by-segment as opposed to averaging over portions of data with different noise characteristics. Several key aspects of our SBI strategy allow us to realise these benefits efficiently in a full analysis pipeline.
\begin{itemize}[leftmargin=*]
\item \textbf{Amortised Marginal Posterior Estimates:} The full dataset is a series of $N_s$ segments, each with independent noise parameters $(A_\tau, P_\tau)$, with $\tau = 1 \ldots N_s$, but shared signal parameters $\theta_s = (\alpha, \gamma)$ which we ultimately want to infer. The independence of the segments allows us to break up the calculation of the signal posterior $p(\alpha, \gamma | \{d_1, \ldots, d_{N_s}\})$ into $N_s$ sub-components:
\begin{align}
&\frac{p(\alpha, \gamma | \{d_1, \ldots d_{N_s}\})}{p(\alpha, \gamma)} \nonumber \\ 
&= \int{\mathrm{d} A_1 \, \mathrm{d} P_1 \cdots \mathrm{d} A_{N_s}} \,\mathrm{d}P_{N_s} \, \nonumber \\ 
&\qquad \qquad \frac{p(\alpha, \gamma, A_1, P_1, \ldots, A_{N_s}, P_{N_s} | \{d_1, \ldots d_{N_s}\})}{p(\alpha, \gamma)} \nonumber \\
&= \int{\mathrm{d} A_1 \, \mathrm{d} P_1 \cdots \mathrm{d} A_{N_s}} \,\mathrm{d}P_{N_s} \, \frac{p(\alpha, \gamma, \ldots, A_{N_s}, P_{N_s})}{p(\alpha, \gamma)} \nonumber \\ 
&\qquad \qquad \times \frac{p(\{d_1, \ldots d_{N_s}\} | \alpha, \gamma, A_1, P_1, \ldots, A_{N_s}, P_{N_s})}{p(\{d_1, \ldots d_{N_s}\})} \nonumber \\
&= \int{\mathrm{d} A_1 \, \mathrm{d} P_1 \cdots \mathrm{d} A_{N_s} \,\mathrm{d}P_{N_s}} \, \frac{p(d_1 | \alpha, \gamma, A_1, P_1)}{p(d_1)} \cdots \nonumber \\ 
&\qquad \frac{p(d_{N_s} 
| \alpha, \gamma, A_{N_s}, P_{N_s})}{p(d_{N_s})} \frac{p(\alpha, \gamma, \ldots, A_{N_s}, P_{N_s})}{p(\alpha, \gamma)}\nonumber \\
&\qquad \quad \times \frac{p(d_1)\cdots p(d_{N_s})}{p(\{d_1, \ldots d_{N_s}\})} \nonumber \\
&\propto\int{\mathrm{d} A_1 \, \mathrm{d} P_1 \cdots \mathrm{d} A_{N_s} \,\mathrm{d}P_{N_s}} \, \frac{p(\alpha, \gamma, A_1, P_1 | d_1)}{p(\alpha, \gamma, A_1, P_1)} \cdots \nonumber \\ 
&\qquad \frac{p(\alpha, \gamma, A_{N_s}, P_{N_s} | d_{N_s})}{p(\alpha, \gamma, A_{N_s}, P_{N_s})} \frac{p(\alpha, \gamma, \ldots, A_{N_s}, P_{N_s})}{p(\alpha, \gamma)}\nonumber \\
&= \frac{p(\alpha, \gamma | d_1)}{p(\alpha, \gamma)} \cdots \frac{p(\alpha, \gamma | d_{N_s})}{p(\alpha, \gamma)} = \prod_{\tau = 1}^{N_s} \frac{p(\alpha, \gamma | d_\tau)}{p(\alpha, \gamma)}. \label{eq:marginal_posts}
\end{align}
where the proportionality factor in the second-to-last line is only a function of the data $\{d_1, \ldots\}$ and in the last line we have assumed that the prior $p(\alpha, \gamma, A_1, \ldots)$ factorises across the segments. Given our assumptions about the noise specified above, we assume that the spectral shapes of the noise components do not vary, only their magnitude as specified by the parameters $A$ and $P$. This means that each of the marginal posteriors $p(\alpha, \gamma | d_\tau)$ are solving the \emph{same} Bayesian inference problem segment-by-segment, albeit with varying realizations of $d_\tau$, drawn from different noise parameters. In the SBI context, this means that we only need to solve the one-segment problem \emph{once}, and then we can apply the trained network to the remaining segments essentially for free, once a well-calibrated marginal posterior estimate $\hat{p}(\alpha, \gamma | d_\tau)$ is obtained. In terms of performance, with a trained network, we can analyze a single segment in $\mathcal{O}$(seconds) and the full dataset in $\mathcal{O}$(1 minute)\footnote{\label{foonote:benchmark}For a more precise benchmark, the computing hardware used in this work consisted of 18 (shared) Intel(R) Xeon(R) Platinum 8360Y CPUs and a single, shared NVIDIA A100 graphics card. With the learning rate scheduler and network described below, training takes $\mathcal{O}$(1 hour).}.
\item \textbf{Low-dimensional Data Representations:} Arguably, one of the key challenges for analysing LISA data in an SBI context is data storage. To train an SBI algorithm requires data $d$ generated according to the implicit data likelihood $p(d | \theta)$ across a range of possible model parameters $\theta$. Breaking down the inference problem as above already helps in this regard because it allows us to store only data for a \emph{single} segment but use it to train a network capable of analysing \emph{all} segments. In addition, because SBI is an implicit likelihood technique, we can actually perform any transformation to the data $d \rightarrow f(d)$ and the set of algorithms can (in principle) learn the corrected likelihood for the compressed or transformed data $p(f(d) | \theta)$. Here, we take a similar approach to~\cite{Flauger:2020qyi,Caprini:2019pxz} and coarse-grain the data across the almost $500,000$ frequency bins\footnote{The $500,000$ frequency bins comes from the following computation: we have segments of length $T = 11.5\,\mathrm{days}$ which means that we have a minimum frequency resolution of $\Delta f = 1/T = 10^{-6} \, \mathrm{Hz}$. Taking a minimum frequency of $f_\mathrm{min} = 3 \times 10^{-5} \, \mathrm{Hz}$ and $f_\mathrm{max} = 5 \times 10^{-1}\,\mathrm{Hz}$, the total number of frequency bins in the $N = (f_\mathrm{max} - f_\mathrm{min})/\Delta f = 499,970$ i.e. about 500,000.}. In particular, we first fix the relevant coarse-grained frequencies before directly averaging the data inside the bins defined by this grid. An example of this coarse-graining process is shown in Fig.~\ref{fig:data}. In the analysis shown below, we coarse-grain the quadratic data down from $\sim 500,000$ frequency bins to $\sim 2000$\footnote{We have also checked explicitly that we can achieve the same posteriors with far fewer bins than this, at least as low as 200. This immediately decreases the training and inference time and reduces data storage requirements.}. This has three key benefits as far as SBI is concerned: firstly, we can retain all the necessary information in the coarse-grained representation (as shown by the MCMC comparison); secondly, the data storage requirements are significantly reduced, since we only need to store coarse-grained data for a single segment; and thirdly, the network training is significantly accelerated, since the architecture only has to process data with $\sim6000$ dimensions, as opposed to the $\sim$1.5 million that one would nominally start with. We describe the network architecture in more detail below, but it is worth noting that this sort of compression on a segment-by-segment basis may not possible in an MCMC context if the exact likelihood for the compressed data $p(f(d) | \theta)$ cannot be written down explicitly\footnote{\label{foonote:data_compression}We note, however, that it is possible to write approximate likelihoods for the compressed data. For example, in~\cite{Flauger:2020qyi} an approximate likelihood is defined for coarse-grained data that are the result of averaging over many time segments. Alternatively, in~\cite{Baghi:2023qnq}, a Wishart likelihood is used to model data averaged across frequency bins.}.
\item \textbf{Efficient Data Generation:} The final aspect that allows us to make our pipeline efficient to train and carry out inference with, is our data generation and storage scheme. Due to the particular setup of this simulation model, the three relevant components (TM noise, OMS noise, and signal) have a known scaling with frequency. We can leverage this and significantly bootstrap our sample size by splitting up simulations into component-by-component blocks. Specifically, we are ultimately interested in analysing the coarse-grained quantity given schematically by $f_\mathrm{CG}(d^{I\star} d^{I} = |n^I_\mathrm{TM} + n^I_\mathrm{OMS} + s^I|^2)$ where $f_\mathrm{CG}$ represents the coarse-graining operator. This operation is linear, so we can write this quantity as:
\begin{multline}
f_\mathrm{CG}(|n^I_\mathrm{TM} + n^I_\mathrm{OMS} + s^I|^2) = f_\mathrm{CG}(|n^I_\mathrm{TM}|^2) \\ + f_\mathrm{CG}(|n^I_\mathrm{OMS}|^2) + f_\mathrm{CG}(|s^I|^2) + f_\mathrm{CG}(2 \mathrm{Re}(n^{I \, \star}_\mathrm{TM} n^I_\mathrm{OMS})) \\ + f_\mathrm{CG}(2 \mathrm{Re}(n^{I \, \star}_\mathrm{OMS} s^I)) + f_\mathrm{CG}(2 \mathrm{Re}(n^{I \, \star}_\mathrm{TM} s^I)) 
\end{multline}
The crucial point is that each of these terms has a known scaling with the parameter values and/or a given frequency dependence through the templates. For example, $f_\mathrm{CG}(|n^I_\mathrm{TM}|^2) \propto A^2$, $f_\mathrm{CG}(2 \mathrm{Re}(n_\mathrm{TM}^{I \, \star} n^I_\mathrm{OMS})) \propto AP$ etc. This means, that if we generate and store these 6 independent coarse-grained quantities, we can generate a ``new" simulation for any combination of parameters (or indeed any signal/noise template, making this a very general purpose approach). Note that these quantities are all expressed in the relevant TDI basis (AET in this paper). Crucially, this ensures that the (relatively) expensive signal response, noise generation, and TDI basis transformations need only be carried out \emph{once}. In practice, we generate $50,000$ realizations of each term in this sum which we can use for all training and analyses\footnote{\label{foonote:data_gen}From a technical perspective, we write a \texttt{pytorch}~\cite{pytorch_2019} dataset and data loader that loads coarse-grained examples and rescales them according to sampled parameter values. This is fast enough to be carried out online, during training. The data generation, paralellised across 10 (shared) Intel(R) Xeon(R) Platinum 8360Y CPUs, took $\mathcal{O}(1-2)$ hours.}. This saves significantly on simulation costs that would otherwise typically be associated with sequential SBI methods, see \emph{e.g.}~\cite{Lueckmann:2017aaa,Miller:2021hys,Miller:2022shs}.
\end{itemize}

\begin{figure}
    \centering
    \includegraphics[width=\linewidth]{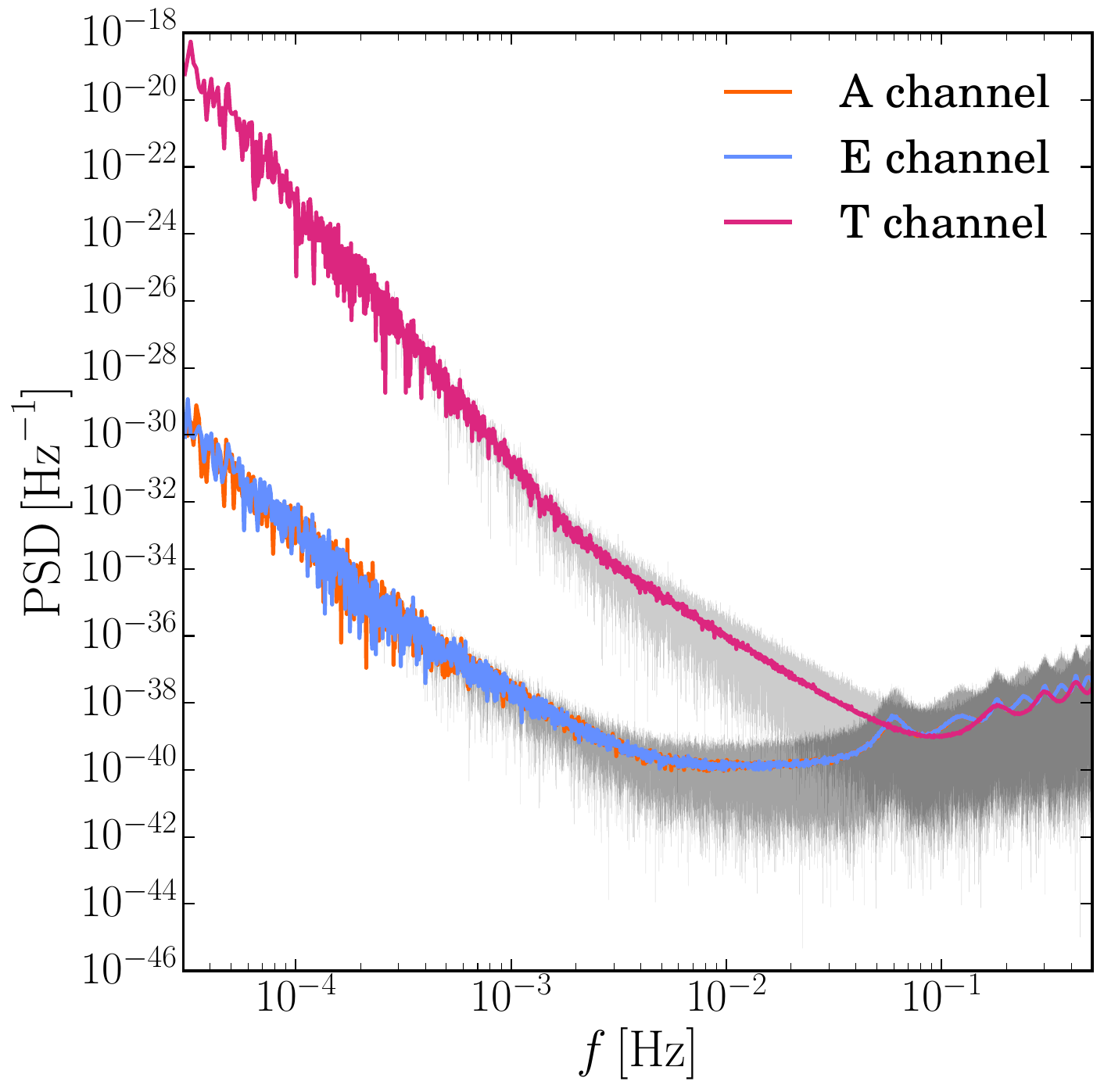}
    \caption{Coarse-grained Data. Realization of the coarse-grained data for the A (orange), E (blue), and T (pink) channels used in the analysis. The black curves in the background are the uncompressed quadratic data across the full frequency grid.}
    \label{fig:data}
\end{figure}

\subsection{SBI Implementation and Technical Details}

\noindent \textbf{SBI Algorithm:} In this work, and the corresponding \texttt{saqqara} code, we use the Truncated Marginal Neural Ratio Estimation (TMNRE) algorithm~\cite{Miller:2021hys,Miller:2022shs}\footnote{Note this is not the only option, and the core aspects of our proposed framework will remain unchanged regardless of SBI algorithm, up to and including the compression aspect of the network design. Other options include Neural Likelihood Estimation (NLE)~\cite{Papamakarios:2016aaa,Alsing:2019xrx} or Neural Posterior Estimation (NPE)~\cite{Papamakarios:2016aaa} and their sequential versions.}. This is described in detail in Refs.~\cite{Alvey:2023pkx,Bhardwaj:2023xph,Miller:2021hys,Miller:2022shs}, see in particular Appendix D in Ref.~\cite{Alvey:2023npw} where the first version of \texttt{saqqara} was released and a detailed discussion surrounding the algorithm and design choices are presented. To recap, however, the key points of the method are:
\begin{itemize}[leftmargin=*]
\item Given simulation data $(x, \theta)$ such that $x \sim p(x | \theta)$, \emph{i.e.}, $x$ is sampled from the data likelihood under $\theta$, one can construct a binary classification problem between joint samples $(x, \theta) \sim p(x, \theta) = p(x | \theta) p(\theta)$ and marginal samples $(x, \theta) \sim p(x) p(\theta)$. Training data for the latter class is easy to generate from simulation data by simply permuting $\theta$ amongst examples or resampling $\theta \sim p(\theta)$. By parametrizing (typically through a neural network) a classifier $f_\phi(x, \theta)$ and optimising the binary cross entropy loss (which is the standard choice for binary classification tasks to assign one of two classes to data examples, see e.g.~\cite{Miller:2022shs} for its use in the context of parameter estimation),
\begin{figure}
    \centering
    \includegraphics[width=\linewidth]{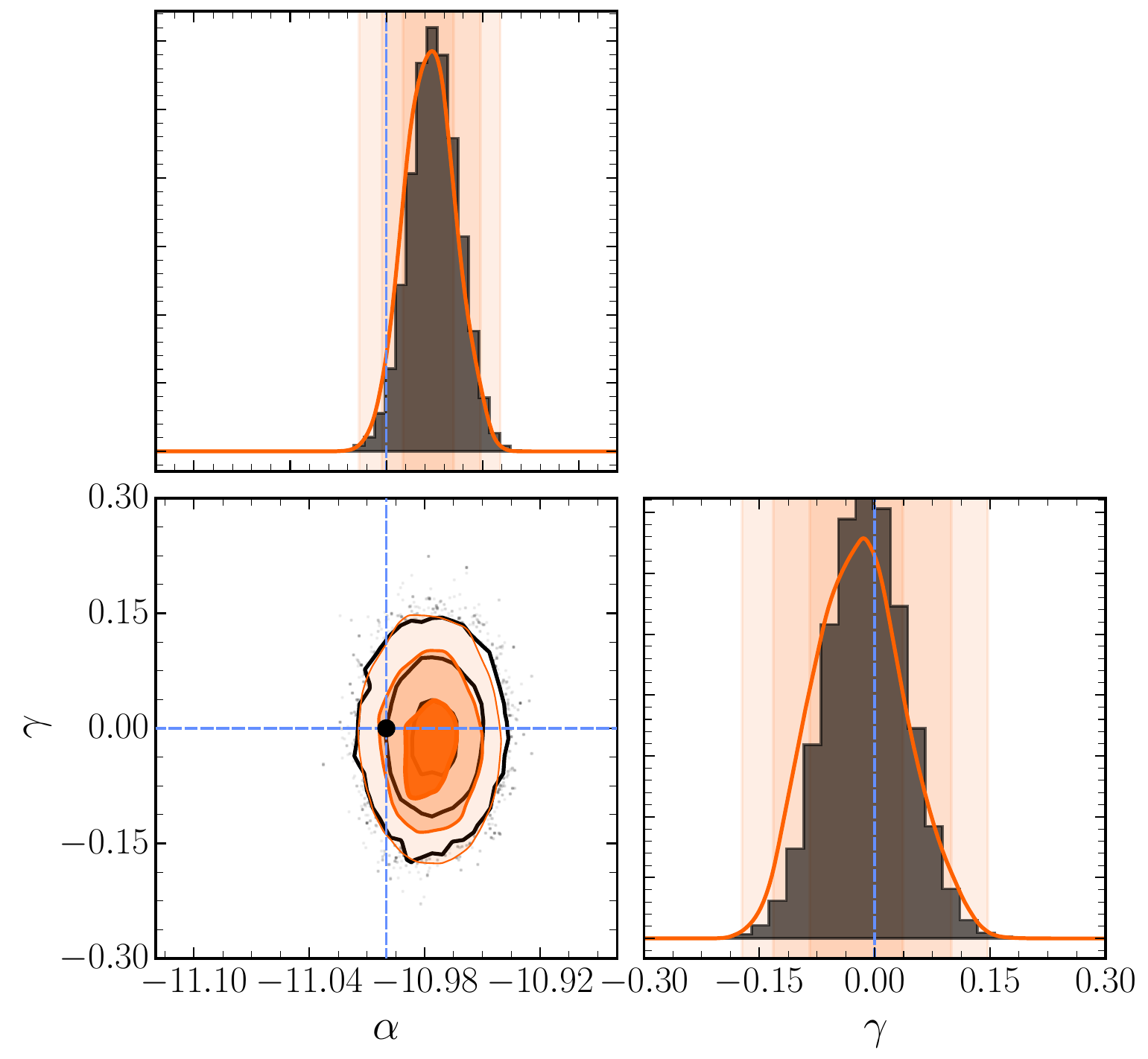}
    \caption{MCMC vs. SBI Comparison. Comparison between MCMC (black contours and histogram) and SBI (orange contours) at the single segment level, evaluated on the same noise realization. The MCMC analysis uses the Whittle likelihood, evaluated on the full frequency grid. The blue lines indicate the injection values for the signal parameters $\alpha$ (amplitude) and $\gamma$ (slope).}
    \label{fig:mcmc_comparison}
\end{figure}
\begin{multline}
\mathcal{L}[f_\phi] = -\int \mathrm{d}x\,\mathrm{d}\theta \, p(x, \theta) \log \sigma(f_\phi(x, \theta)) \\ + p(x)p(\theta) \log \sigma(-f_\phi(x, \theta)),
\end{multline}
with respect to $\phi$ (the trainable parameters of the neural network $f_\phi$), it can be shown that the optimal classifier is given by $f_\phi^\star(x, \theta) = \log \left[ p(x, \theta)/p(x)p(\theta) \right] = \log \left[ p(\theta | x) / p(\theta) \right]$. In the expression above, $\sigma(x) = (1 + \exp(-x))^{-1}$ is the sigmoid function. In other words, we can directly obtain the posterior-to-prior ratio by optimising this loss. For low-dimensional problems, this can be evaluated on a suitably dense grid of $\theta$, or it can be efficiently sampled, see, \emph{e.g.}, Ref.~\cite{AnauMontel:2023stj}. If an alternative such as Neural Posterior Estimation~\cite{Papamakarios:2016aaa} was used instead, then the density estimator could be directly sampled or evaluated.
\item Importantly, this classification problem can be carried out on either the full parameter vector $\theta$ or any subset of $\theta$ to obtain the full joint posterior or any desired marginal posterior. We use this feature here to target specifically the marginal posteriors for the signal parameters $p(\alpha, \gamma | d_\tau)$, motivated by the discussion above. This allows us to combine the information from segment to segment immediately at the level of the marginal posteriors rather than having to marginalise over the noise parameters of each segment explicitly.
\end{itemize}

\noindent \textbf{Network Architecture:} Our algorithm is designed to analyze coarse-grained, quadratic data in the frequency domain across three data channels (here AET, but this is not a strict choice). We designed our network architecture accordingly, with the knowledge that: the normalization across the channels differs in the raw data (\emph{e.g.}, the TT channel is dominated by OMS noise), the data can vary by orders of magnitude across the frequency grid, and the information about spectral shapes is contained in data spanning the full frequency grid.

With these facts in mind, given coarse-grained quadratic data $D_\tau^I = f_\mathrm{CG}(d_\tau^{I \star}d_\tau^I)$ the first step is to normalise across the dataset both $\log D_\tau^I$ and $D_\tau^I$ separately, channel-by-channel and frequency-by-frequency, \emph{i.e.}, we essentially whiten the data independently across A, E, T, and frequency. Then, two copies of a standard \texttt{resnet} architecture with $3$ input channels, $64$ hidden features, and $2$ blocks are applied to the normalised versions of both $\log D_\tau^I$ and $D_\tau^I$. These networks output a compressed version of the data with shape (3, \# of parameters to infer -- which here is two, $\alpha$ and $\gamma$). More broadly, the aim of the \texttt{resnet} part of the architecture is to summarise the data into a form (and lower dimensionality) where the binary classification task is simpler to carry out. This is a standard aspect in many SBI algorithms and outputs a ``learned summary statistic", which is in some sense optimised for the classification task. We note that, of course, it is possible to work with some pre-constructed summary, but for complex datasets/simulators, this hand-crafted summary may lose information, so the ability to simultaneously construct a data summary as part of the algorithm is a useful feature. The second part of the network takes this learned summary statistic and uses the standard ratio estimator implemented in the \texttt{swyft} code~\cite{Miller:2021hys,Miller:2022shs} to estimate the 2-dimensional ratio $p(\alpha, \gamma | s_\phi(D_\tau^I)) / p(\alpha, \gamma)$ where $s_\phi(D_\tau^I)$ is the learned summary statistic of the input data $D_\tau^I$. Importantly, both parts of the network are trained simultaneously under the binary cross-entropy loss defined above to obtain a single estimator $f_\phi(D_\tau^I, (\alpha, \gamma)) = \tilde{f}_\phi(s_\phi(D_\tau^I), (\alpha, \gamma))$ which first compresses the data, then computes the marginal posterior. We have tested that moderate changes to this architecture, \emph{e.g.}, increasing the number of residual blocks, or changing the number of hidden features have no impact on the inference quality. 

\vspace{6pt}
\noindent \textbf{Training Details:} As discussed, we train the classifier according to the binary cross entropy loss. The parameters of the network are optimised using the \texttt{Adam} optimiser with a learning rate scheduler that consists of an initial 20 step linear warm up from zero learning rate to a maximum learning rate of $5 \times 10^{-5}$. The learning rate then decays with a cosine scheduler for a total of $380$ steps to a minimum learning rate of $8 \times 10^{-7}$. Given that this learning rate schedule is somewhat conservative, we expect that the number of steps (and therefore the training time) in this cosine decay step can be reduced. In total, we use $50,000$ simulations per epoch split into 70\% training steps, and 30\% validation steps. We use a relatively large batch size of $8192$ for training and validation, although as with the network architecture, we found that moderate changes to this setup, \emph{e.g.}, up to maximum learning rates of $\sim 1 \times 10^{-4}$ or down to the batch size of $\sim 1024$ had no impact on the inference results. The training and validation curves as well as the learning rate schedule for the production run can be found in the Appendix.

\vspace{6pt}
\noindent \textbf{Prior Truncation:} For the case study at hand, the parameter reconstruction is extremely precise, and the posterior is highly peaked compared to the initial prior we take. To ensure that we are efficiently accessing the information, we use a prior truncation scheme to initially ``zoom-in" on the relevant region of signal parameter space. This follows the approach described in Refs.~\cite{Bhardwaj:2023xph,Alvey:2023npw}, where the inference proceeds in rounds, and regions of extremely low posterior(-to-prior) are removed from the prior and the data re-simulated. Given our efficient data simulation technique described above, this does not require any further simulation runs, only a change in the prior evaluation at training time. For the production run provided here, the inference proceeds in two rounds, with an initial signal prior given by $p(\alpha, \gamma) = \mathrm{U}_\alpha(-13, -8) \times \mathrm{U}_\gamma(-5, 5)$. In the future, we plan also to explore online active learning techniques to avoid the need for discrete rounds (see \emph{e.g.}~\cite{Gloeckler:2024aaa}). We will also explore techniques for zooming in to relevant regions of noise parameter space across full mission duration data. 

\subsection{Results}

\noindent Following the training and implementation details described above, we can now test our SBI framework. First, we ascertain whether we match the expected posteriors within a single segment. This is to ensure that when we combine segments, following Eq.~\eqref{eq:marginal_posts}, we obtain unbiased and accurate results. To carry out this test, we implement the full Whittle likelihood across all frequency bins and channels (\emph{i.e.}, written in terms of the non coarse-grained data, as discussed above) provided in App.~\ref{app:fisher}. We then use the \texttt{emcee} sampler~\cite{Foreman-Mackey:2012any} to evolve 100 walkers for 1000 steps. The comparison to the SBI approach is shown in Fig.~\ref{fig:mcmc_comparison} for the fiducial parameters $(\alpha, \gamma, A, P) = (-11, 0, 3, 15)$. We see that we achieve excellent agreement in this very high precision setting with the classical sampling approach. In addition to this single observation test, we also show the expected vs. empirical coverage of our trained network in the right hand panel of Fig.~\ref{fig:training}.

\begin{figure}
    \centering
    \includegraphics[width=\linewidth]{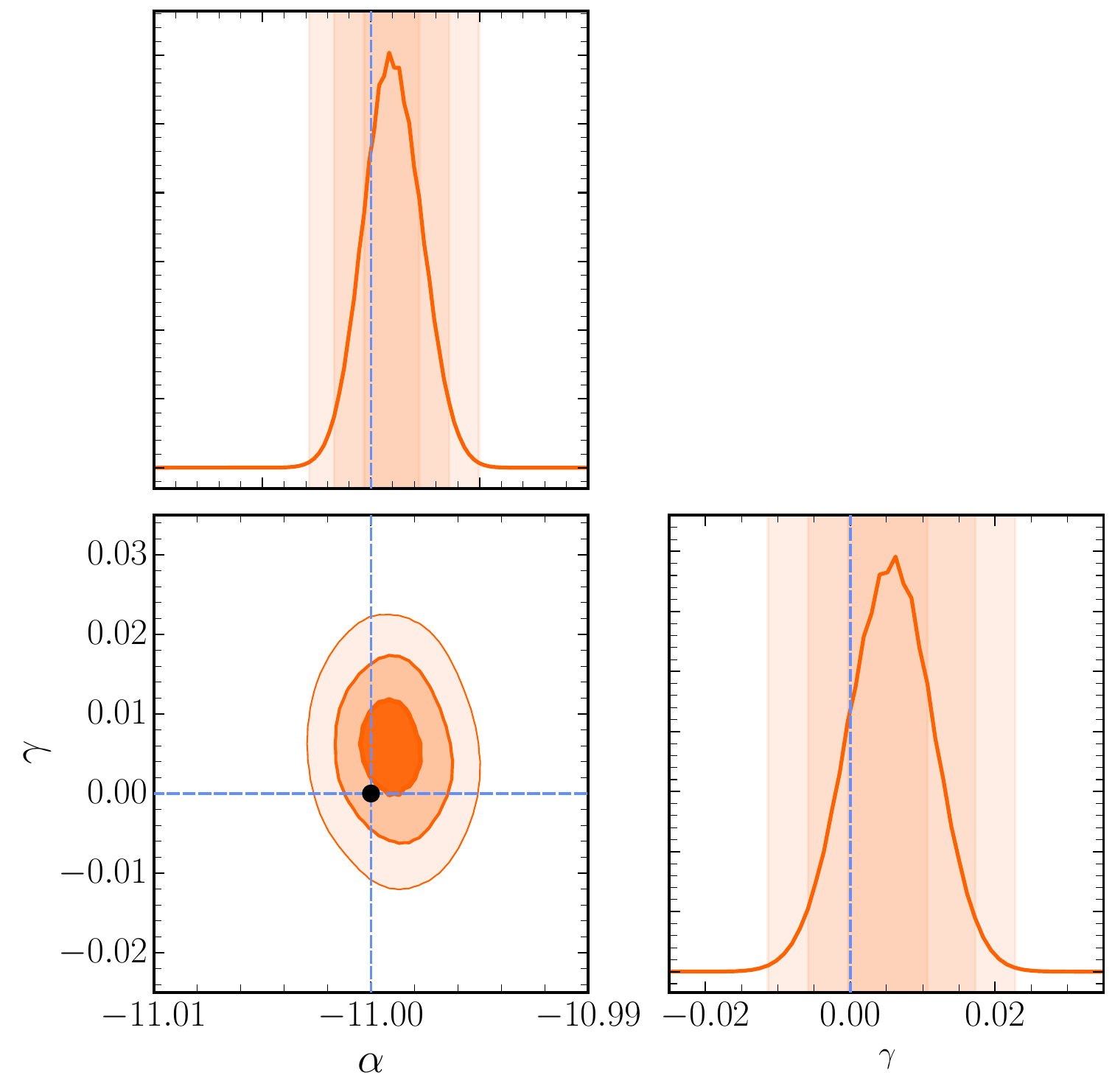} 
    \caption{Full mission duration inference. Example of a combined inference result across $N_s = 100$ segments with the SBI method described in this work. The blue lines indicate the true injection values.}
    \label{fig:full_inference}
\end{figure}

With this agreement established, we can now look to combine segments. To do so efficiently, for a given observation we evaluate our amortized (log)-ratio estimator $\log[\hat{p}(\alpha, \gamma | D_\tau) / p(\alpha, \gamma)]$ on a (2000, 2000) grid of $\alpha, \gamma$ values. These can then be used to compute the total, un-normalised (log)-posterior $\log p(\alpha, \gamma | \{D_1, \ldots, D_{N_s}\}) = \sum_\tau \log[\hat{p}(\alpha, \gamma | D_\tau) / p(\alpha, \gamma)] + \mathrm{const.}$. We found this grid strategy led to significantly more stable posteriors once combined compared to a random sampling of the prior space. An example of the resulting posteriors is shown in Fig.~\ref{fig:full_inference}. We emphasize the unbiased nature of the posterior, which is non-trivial to achieve in this context. Indeed, even slight inaccuracies in the single segment results, when combined for many segments, can lead to biased results. This feature demonstrates the high level of accuracy we have achieved in the single-segment results.

\begin{figure}
    \centering
    \includegraphics[width=\linewidth]{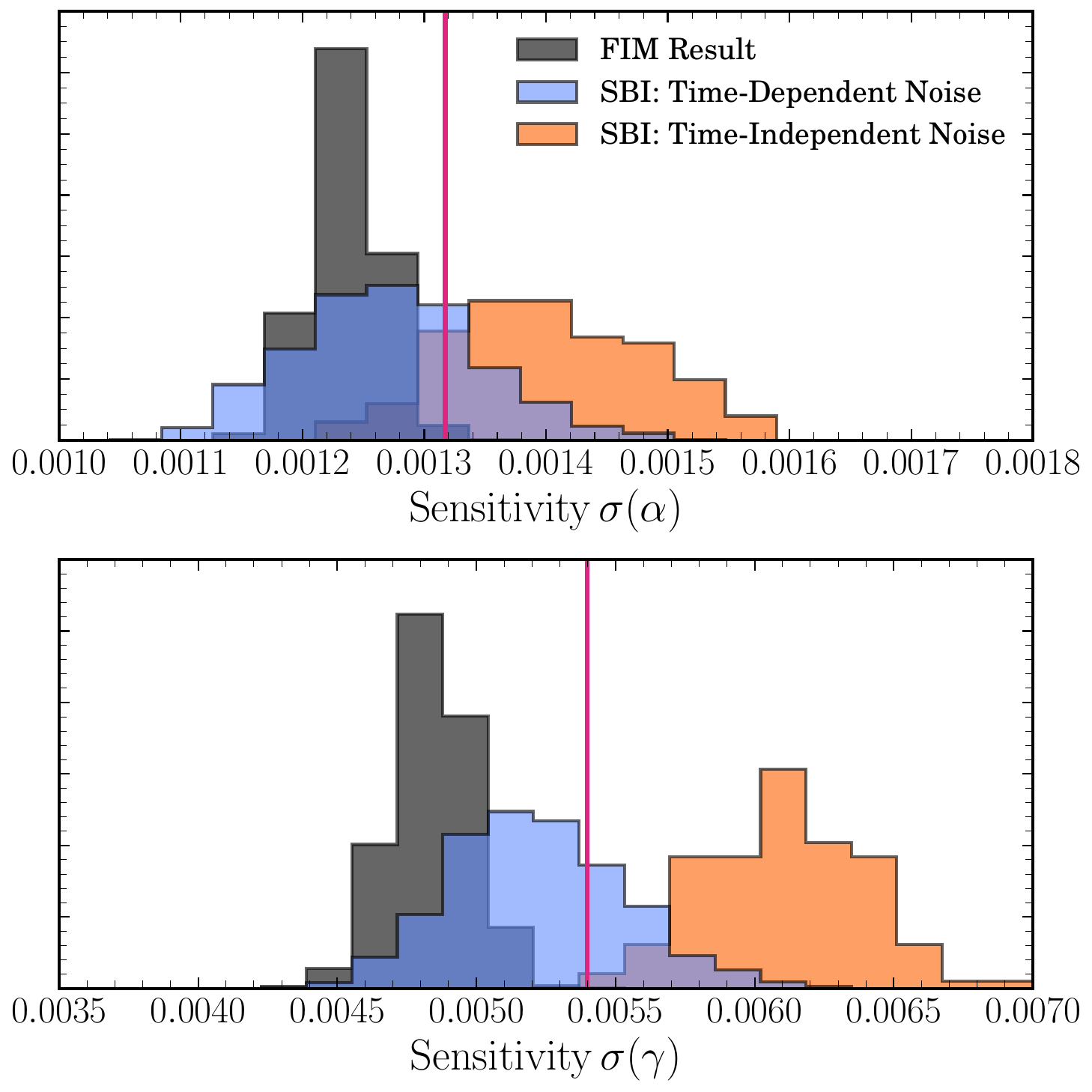} 
    \caption{Main result of this work. Sensitivity results for the $\alpha$ (amplitude) and $\gamma$ (slope) parameters. The blue histogram shows the results of analysing data with varying instrumental noise, orange instead for fixed noise parameters across all the segments. The black histograms and the pink vertical lines are the results of the FIM approach for comparison. We see that we can obtain the expected sensitivity improvement with our SBI method. Note that there is some additional spread compared to the FIM method as a result of noise fluctuations when analyzing the full data.}
    \label{fig:sbi_results}
\end{figure}

Finally, the key results of this work are shown in Fig.~\ref{fig:sbi_results}. Here, we show the resulting sensitivity for reconstructing the signal parameters using our SBI method. The black histogram and pink vertical line represent the FIM results derived in the last section for comparison. The blue histogram is the result of using our SBI pipeline on realizations of a full mission dataset consisting of $N_s = 100$ segments, each with instrumental noise that has different underlying values of $A$ and $P$, sampled as described in Sec.~\ref{sec:modeling}. The orange histogram then shows the sensitivity obtained analysing data generated with noise parameters fixed to their fiducial (mean) values $A = 3$, $P = 15$. We see that we indeed obtain the predicted sensitivity increase found in Sec.~\ref{sec:fisher} when analysing the varying noise case.

\subsection{Generalisability and Pros/Cons vs. Traditional Methods}

\noindent So far we have discussed the strategy that can be employed to efficiently analyze LISA data when time varying instrumental noise is present. This broadly revolved around breaking the analysis into segments and developing an amortised analysis pipeline that can be applied to each segment before combining. We showed that we could achieve the estimated performance improvement in the signal reconstruction conjectured with the FIM analysis. We also showed that we can achieve the necessary robust agreement with the high precision MCMC result on a segment-by-segment basis. As we look forward to applying this to increasingly realistic scenarios, it is important to evaluate the generality of the approach as well as the relative pros and cons versus, \emph{e.g.}, a traditional approach.

\vspace{6pt}
\noindent \textbf{Generalisation of strategy:} The general strategy employed in this work is to analyze cosmological SGWBs in LISA data segment-by-segment using a pre-trained network to combine individual segments into one informative posterior. Looking forward, we are interested as to how well this strategy can generalise to more complex scenarios. To do so, we identify some challenges and possible solutions. Firstly, the assumption regarding the independence of individual segments could be broken. This would happen for instance if working with data containing long-lived transient signals that exceed segment boundaries, see Sec.~\ref{sec:signal_model} for a more detailed discussion and possible mitigation strategies. Secondly, realistic signal and noise models will likely be more complicated than the simple templates used here. As long as these can be described by parametric models, \emph{i.e.}, they allow for forward modeling, the tools developed here can be straightforwardly generalized. In fact, the modular structure of our code is set up to facilitate this. On the signal side, this may include complicated analytical templates or template data banks. On the noise side, this may involve noise models with many different noise components, whose parameters may vary independently. Moving towards more agnostic models implies implementing more flexible templates, typically with a parameter number that varies dynamically while performing the inference. Exploring such scenarios requires some mechanism for model selection (such as reversible jumps in MCMC) to be in-built into the inference pipeline. In principle, SBI already directly allows for model comparison~\cite{Jeffrey:2023stk}. However, tackling this issue would correspond to simulating data and training a network over several parameter spaces simultaneously. While, conceptually, there are no clear no-goes in performing this procedure, technically, some parts of our training algorithm (mostly related to truncation) would have to be adapted to face this challenge. We plan to return to this in future work. In this work, we have also assumed that the changes in noise parameters occur at discrete times. In reality, it may be the case for some noise components that the change is more adiabatic, occurring slowly over a segment. It is possible to include this directly in the framework just by allowing the noise parameters to vary in time during a segment and then generate the data accordingly. This is because SBI can still learn to marginalize over this non-stationary noise, and the benefits surrounding the fact there is one stationary cosmological component remain. Indeed, one such example of this could be the background of galactic binaries which vary over the year. In this case we could additionally think about conditioning the trained estimator directly on the time at which the segment starts $\tau$, allowing us to leverage our knowledge of the annual modulation, but maintain the benefits of only having to train one network.  Finally, there is the challenge of data generation and storage. Specifically, to maintain the efficient data generation step described above, we needed to store both the coarse-grained, squared data as well as the cross-terms (totalling 6 different pieces of data per simulation). As the number of modeling components increases, constructing and storing the data this way introduces combinatorially increasing storage requirements. This motivates the further investigation of linear data compression steps, or indeed analysis carried out directly at the linear level, which would then only scale linearly in the number of components. This would also allow us to trivially move beyond the (implicit) assumption of a diagonal noise basis, which here is the AET basis. We will explore these avenues in future work and releases of \texttt{saqqara}.

\vspace{6pt}
\noindent \textbf{Strengths of SBI Approach:} Beyond the applicability of the methods to more complex setups, there are several benefits compared to traditional methods. The first is the inherent ability of SBI to analyze bespoke/flexible data representations. Here, we use that to our advantage by analysing a coarse-grained version of the data. This speeds up inference substantially (by a factor of approximately $10^3$ here given by the ratio of the number of initial frequency points to the number of coarse-grained bins), and reduces storage requirements. At the segment-by-segment level, this compression may not be possible with traditional sampling approaches, since it is not possible to explicitly write down the exact likelihood for the coarse-grained data (see footnote~\ref{foonote:data_compression}). The second major benefit is the ability to amortise the analysis over segments. In particular, by training the network to analyze one segment, we then get the posteriors for the other segments ``for free". In a sequential sampling-based approach, however, each segment must be analyzed on the full frequency grid from scratch each time. For $N_s = 100$ segments, this can lead to a significant inference cost given a single MCMC run on one uncompressed data segment takes $\mathcal{O}(1)$ hour\footnote{As mentioned above in footnotes~\ref{foonote:benchmark} and~\ref{foonote:data_gen}, we are running this on hardware consisting of 18
(shared) Intel(R) Xeon(R) Platinum 8360Y CPUs.}. In comparison, for each set of $N_s = 100$ segments, given a trained network, we can analyze the full mission dataset in $\mathcal{O}(1)$ minute. A final advantage of the SBI approach is that we can directly target the inference results we are most interested in. For example, using TMNRE (although it is possible with other SBI methods) we can directly access the posterior for the signal parameters, marginalised over the noise model. The important point to note here is that the scaling of this method is in principle independent of the parametric complexity of the noise model, whereas classical sampling run-times must scale with the full dimensionality of the problem.

\vspace{6pt}
\noindent \textbf{Open Challenges for SBI Approach:} Currently, the main challenges facing SBI surround convergence checks and the identification of mismodeling~\cite{AnauMontel:2024flo}. Specifically, in the SBI context, we are often in the position where inference results are optimal conditional on simulation dataset size, and network architecture. In other words, for a given problem, it is possible that an increase in training dataset size, or a change in network architecture could improve the sensitivity of the method. This is compared to convergence tests such as the Gelman-Rubin criteria for MCMC. In this work here, of course, we made direct comparisons so could benchmark our performance with Fisher analyses and MCMC runs, however, more generally this motivates the further development of these tools for SBI applications. On a related note, some of the same convergence guarantees are not available for SBI techniques - such as the long-term Markov Chain behaviour of MCMC. As such, there is a certain reliance on stable network training and faithful data representations to obtain accurate posteriors. On the other hand, additional cross checks, such as testing coverage, can be a complementary diagnostic for an analysis pipeline. Finally, as the modeling of LISA data becomes more complex and detailed, the identification of systematic effects and any mismodeling, either via the likelihood or the simulator in the SBI context is a crucial step. Again, we would argue that this need motivates the more general development of goodness-of-fit tools and checks as SBI becomes a more widespread analysis technique. In addition, it underlines the need for several different analysis chains, based on different methods and making different modeling choices.

\vspace{-8pt}
\section{Conclusions and Outlook}\label{sec:conclusions}

\noindent The key result of this work is to establish that an analysis of LISA data fragmented into time segments can leverage variations in the instrument noise to increase the sensitivity to a stochastic gravitational wave background. Such noise variations are expected to occur after scheduled satellite operations or unscheduled glitches. However, fragmenting the data into time segments poses a challenge for traditional data analysis techniques in terms of efficiency. We demonstrate that SBI techniques are well adapted to take up this challenge: The network can be trained and amortised on a single data segment, using frequency coarse-grained data. Obtaining the full inference result then simply amounts to combining the marginal posterior estimates of the different data segments. Marginalization over noise variations is inbuilt, returning directly the posterior estimates for the signal parameters of the SGWB. The successful implementation of this procedure is illustrated in Fig.~\ref{fig:sbi_results}, which shows the sensitivity increase achieved by performing a segment-by-segment analysis with our SBI pipeline, together with estimates of the expected effect using FIM forecasts. 

This work should be seen as proof of principle demonstration, relying on simplified assumptions on the noise, signal, and instrument modeling. In particular, we have taken all data segments to have a fixed length of 11.5 days, corresponding to the scheduled reorientation of the satellites. In reality, glitches may occur unscheduled and more frequent than this. On the signal side, we have included only an SGWB and have assumed this signal to be static. Finally, we emphasize that we have used simplified noise and signal templates throughout this work. A generalization to more complex templates is in principle straightforward (see also~\cite{Alvey:2023npw,Dimitriou:2023knw}) and we do not expect this to impact the key results of this work. Going beyond template-based approaches to more agnostic modeling is more challenging, and we plan to return to this in future work.

Our code to simulate and analyze LISA data is publicly available at \href{https://github.com/Mauropieroni/GW_response}{\texttt{GW\textunderscore response}} and \href{https://github.com/peregrine-gw/saqqara}{\texttt{peregrine-gw/saqqara}}. Our simulator features a modular architecture, allowing for independent changes to the signal components, the noise model, and the LISA response functions. We welcome contributions to and the use of this framework aiming to extend the analysis pipelines to a broader class of scenarios. 

In this spirit, we plan to address in future work the inclusion of foreground sources and their parameter reconstruction. Another target is the efficient use of SBI to discriminate SGWB from noise components, based on exploiting differences in the instrument response (\emph{i.e.}, transfer functions) with only minimal input on noise and signal knowledge.

\vspace*{-12pt}
\section*{Acknowledgements}
\noindent JA and MP thank Carlo Contaldi and Olaf Hartwig for the very useful discussion on the subtleties of data generation. This work is part of the project CORTEX (NWA.1160.18.316) of the research programme NWA-ORC which is (partly) financed by the Dutch Research Council (NWO).
Additionally, JA and CW acknowledge funding from the European Research Council (ERC) under the European Union’s Horizon 2020 research and innovation programme (Grant agreement No. 864035). UB is supported through the CORTEX project of the NWA-ORC with project number NWA.1160.18.316 which is partly financed by the Dutch Research Council (NWO). JA acknowledges the hospitality of CERN TH, where some of this work was carried out. JA and MP acknowledge the hospitality of Imperial College London, which provided office space during some parts of this project.  

\bibliography{biblio}

\newpage
\clearpage
\onecolumngrid
\appendix

\section{Derivation of the Fisher Information Matrix}\label{app:fisher}

\noindent In this appendix, we provide a derivation of the Fisher Information Matrix (FIM) used to carry out the analysis in Sec.~\ref{sec:fisher}. Following the notation set up in that section, we start by writing down the (log)-likelihood $\log p_\tau(\{d_{i,\tau}^I\} | \theta_{n,\tau}, \theta_h)$ of frequency-domain strain data $d_{i,\tau}^I = d_\tau^I(f_i)$ in the time segment $\tau = 1...N_s$ for the TDI channel $I$ across frequencies $f_i$, $i = 1...N_k$ (\emph{i.e.}, we work with positive frequencies only) given signal parameters $\theta_h$ and noise parameters (within segment $\tau$) $\theta_{n,\tau}$. Since under the assumptions discussed in Sec.~\ref{sec:TDI}, the AET basis is diagonal, we directly work in this basis, and the total log-likelihood is simply a sum over the three channels. Similarly, we assume no correlations between different frequency bins. Finally, assuming the linear data in channel $I$ to be Gaussian with a variance given by the sum of the instrumental noise PSD $S^I_{n}(f, \theta_{n,\tau})$ and SGWB contribution $S^I_{h}(f, \theta_h)$, in that channel, the (log)-likelihood reads

\begin{figure}[b]
    \centering
    \includegraphics[width=\linewidth]{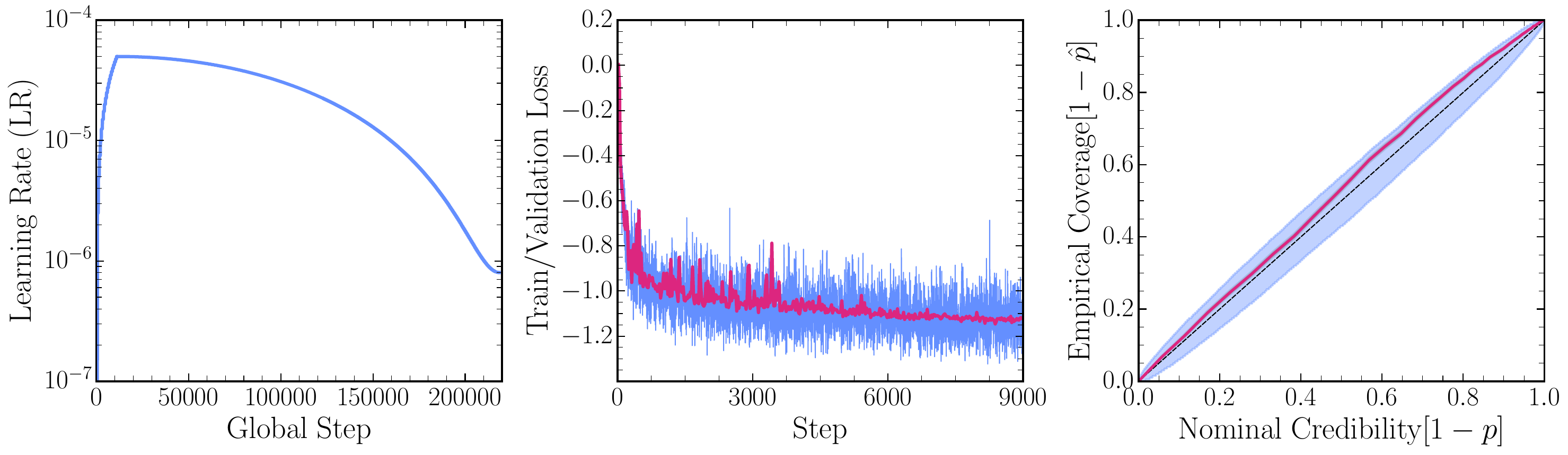}
    \caption{\emph{Left:} Learning rate schedule used for training. In this work we use a linear warm up of 20 steps, followed by a cosine decay. \emph{Middle:} Training (blue) and validation (pink) curves for the production network trained in this work. \emph{Right:} Coverage test results for the single segment analysis. The blue contour shows the 95\% confidence region for the uniform coverage of $4000$ samples. The pink line indicates the observed coverage across the final prior range for our trained network.}
    \label{fig:training}
\end{figure}

\begin{equation}
    \log p_\tau (\{d_{i,\tau}^I\} | \theta_{n,\tau}, \theta_h) = - \sum_{I = \{ {\rm A, E, T}\} } \sum_{i = 1}^{N_k} \left(\frac{d_{i,\tau}^{I\star} \, d_{i,\tau}^I}{S^I_{n}(f_i, \theta_{n,\tau}) + S^I_{h}(f_i, \theta_h)} + \log \left[2\pi (S^I_{n}(f_i, \theta_{n,\tau}) + S^I_{h}(f_i, \theta_h))\right]\right) \; . 
\end{equation}
To simplify the notation, we define the total variance in channel $I$ at a frequency $f_i$ to be $C^I_{i,\tau}(\theta_\tau) = S^I_{n}(f_i, \theta_{n,\tau}) + S^I_{h}(f_i, \theta_h)$, where $\theta_\tau \equiv (\theta_{n,\tau}, \theta_h)$. We can then compute the relevant second derivatives for computing the FIM $F_{\alpha\beta, \tau}$ in time segment $\tau$ via $\partial^2 (-\log p_\tau) / \partial \theta^\alpha_{\tau} \partial \theta^\beta_{\tau} $. These are given by,
\begin{equation}
    \frac{\partial^2 (-\log p_\tau)}{\partial \theta^\alpha_{\tau} \partial \theta^\beta_{\tau} } = \sum_{I = \{ {\rm A, E, T}\} } \sum_{i = 1}^{N_k} \left(\frac{\partial C^I_{i,\tau}(\theta_\tau)}{\partial \theta^\alpha_{\tau}}\frac{\partial C^I_{i,\tau}(\theta_\tau)}{\partial \theta^\beta_{\tau} } \left[-\frac{1}{C^I_{i,\tau}(\theta_\tau)^2} + \frac{2 d_{i,\tau}^{I\star} \, d_{i,\tau}^I}{C^I_{i,\tau}(\theta_\tau)^3}\right] + \frac{\partial^2 C^I_{i,\tau}(\theta_\tau)}{\partial \theta^\alpha_{\tau} \partial \theta^\beta_{\tau} } \left[\frac{1}{C^I_{i,\tau}(\theta_\tau)} - \frac{d_{i,\tau}^{I\star} \, d_{i,\tau}^I}{C^I_{i,\tau}(\theta_\tau)^2}\right] \right).
\end{equation}
The FIM $F_{\alpha \beta, \tau}$ in the segment $\tau$ can then be computed by taking the expectation value $\mathbb{E}$ of this quantity over $\{d_{i,\tau}^I\} \sim p(\{d_{i,\tau}^I\} | \theta_\tau)$. This can be done analytically by noting that $\mathbb{E}_{\{d_{i,\tau}^I\} \sim p(\{d_{i,\tau}^I\} | \theta_\tau)} [d_{i,\tau}^{I \, *} d_{i,\tau}^I] = C_{i,\tau}^I (\theta_\tau)$.

\begin{equation}
    F_{\alpha\beta, \tau}(\theta_\tau) = \mathbb{E}_{\{d_{i,\tau}^I\} \sim p(\{d_{i,\tau}^I\} | \theta^\tau)}\left[\frac{\partial^2 (-\log p_\tau(\{d_{i,\tau}^I\} | \theta^\tau))}{\partial \theta^\alpha_{\tau} \partial \theta^\beta_{\tau} }\right] = \sum_{I = \{ {\rm A, E, T}\} } \sum_{i = 1}^{N_k}\left(\frac{1}{C^I_{i,\tau}(\theta_\tau)^2} \frac{\partial C^I_{i,\tau}(\theta_\tau)}{\partial \theta^\alpha_{\tau}}\frac{\partial C^I_{i,\tau}(\theta_\tau)}{\partial \theta^\beta_{\tau} }\right).
\end{equation}
Finally, taking the continuum limit we approximate $F_{\alpha\beta, \tau}$ as,
\begin{equation}
    F_{\alpha\beta, \tau}(\theta_{n,\tau}, \theta_h) = T_s \sum_{I = \{ {\rm A, E, T}\} } \int_0^\infty  \mathrm{d}f \, \frac{1}{ C^I_{i,\tau}(\theta_\tau)^2} \, \frac{\partial C^I_{i,\tau}(\theta_\tau) }{\partial \theta^\alpha_{\tau}}  \, \frac{\partial C^I_{i,\tau}(\theta_\tau)}{\partial \theta^\alpha_{\tau}}  ,
\end{equation}
where $T_s$ ($\simeq$ 11.5 days in our analysis) is the length of a single time segment. Assuming the segments to be statistically independent, we can then use the FIMs for each segment $\tau$ to construct the overall FIM for the full set of data across segments $\tau = 1...N_s$ (where we take $N_s = 100$ segments above, or about $3$ years of effective data). In particular, if we define the full parameter vector $\theta \equiv (\theta_h, \theta_{n,1}, \cdots, \theta_{n, N_s})$ to consist of the signal parameters $\theta_h$ which are assumed to not vary across segments, and noise parameters $\theta_{n,\tau}$ in each segment $\tau$, then we can construct the non-zero components of the full FIM $F_{\theta_A \theta_B}(\theta)$ (where $A, B$ run over the dimensions of $\theta$). The construction and evaluation of this full FIM $F_{AB}(\theta)$ is implemented following this appendix within the \texttt{sgwbfish} module within the \texttt{saqqara} code.

\end{document}